\documentclass[a4paper,10pt]{article}
\usepackage[pdftex]{graphicx,hyperref}
\usepackage{amsmath,amsfonts,amssymb}
\usepackage{fullpage}
\usepackage{authblk}
\usepackage{setspace}
\usepackage{subcaption}
\usepackage{sidecap}
\usepackage{pbox}
\usepackage{cite}
\usepackage{longtable}
\usepackage[capitalise]{cleveref}
\usepackage{tabularx}
\usepackage{color}
\usepackage{placeins}

\makeatletter
\AtBeginDocument{%
  \expandafter\renewcommand\expandafter\subsection\expandafter
    {\expandafter\@fb@secFB\subsection}%
  \newcommand\@fb@secFB{\FloatBarrier
    \gdef\@fb@afterHHook{\@fb@topbarrier \gdef\@fb@afterHHook{}}}%
  \g@addto@macro\@afterheading{\@fb@afterHHook}%
  \gdef\@fb@afterHHook{}%
}
\makeatother

\newcommand{\beginsupplement}{%
        \setcounter{table}{0}
        \renewcommand{\thetable}{A\arabic{table}}%
        \setcounter{figure}{0}
        \renewcommand{\thefigure}{A\arabic{figure}}%
		\renewcommand{\thesection}{A\arabic{section}}
		\renewcommand{\thesubsection}{A\arabic{subsection}}

     }

\title{\bf{Anticipating cryptocurrency prices \\ using machine learning}}

\author[a]{Laura Alessandretti}
\author[b]{Abeer ElBahrawy}
\author[c]{Luca Maria Aiello}
\author[b,d,*]{\\Andrea Baronchelli}

\affil[a]{{\small Technical University of Denmark, DK-2800 Kgs. Lyngby, Denmark}}
\affil[b]{{\small City, University of London, Department of Mathematics, London EC1V 0HB, UK }}
\affil[c]{{\small Nokia Bell Labs, Cambridge CB3 0FA, UK}}
\affil[d]{{\small UCL Centre for Blockchain Technologies, University College London, UK}}
\affil[*]{{\small Corresponding author:  Andrea.Baronchelli.1@city.ac.uk}}

\date{\today}

\begin{document}
\maketitle

\begin{abstract}
Machine learning and AI-assisted trading have attracted growing interest for the past few years. Here, we use this approach to test the hypothesis that the inefficiency of the cryptocurrency market can be exploited to generate abnormal profits. We analyse daily data for $1,681$ cryptocurrencies for the period between Nov. 2015 and Apr. 2018. We show that simple trading strategies assisted by state-of-the-art machine learning algorithms outperform standard benchmarks. Our results show that non-trivial, but ultimately simple, algorithmic mechanisms can help anticipate the short-term evolution of the cryptocurrency market. 
\end{abstract}

\section*{Introduction}

The popularity of cryptocurrencies has skyrocketed in 2017 due to several consecutive months of super-exponential growth of their market capitalisation~\cite{ElBahrawy170623}, which peaked at more than $\$800$ billions in Jan. 2018. Today, there are more than $1,500$ actively traded cryptocurrencies. Between $2.9$ and $5.8$ millions of private as well as institutional investors are in the different transaction networks, according to a recent survey~\cite{camb2017}, and access to the market has become easier over time. Major cryptocurrencies can be bought using fiat currency in a number of online exchanges (e.g., Binance~\cite{Binance}, Upbit~\cite{Upbit}, Kraken~\cite{Kraken}, etc) and then be used in their turn to buy less popular cryptocurrencies. The volume of daily exchanges is currently superior to $\$15$ billions. Since 2017, over $170$ hedge funds specialised in cryptocurrencies have emerged and bitcoin futures have been launched to address institutional demand for trading and hedging Bitcoin~\cite{sexbitcoin}. 

The market is diverse and provides investors with many different products. Just to mention a few, Bitcoin was expressly designed as a medium of exchange~\cite{nakamoto2008bitcoin,ali2014economics}; Dash offers improved services on top of Bitcoin's feature set, including instantaneous and private transactions \cite{Dash};  Ethereum is a public, blockchain-based distributed computing platform featuring smart contract (scripting) functionality, and Ether is a cryptocurrency whose blockchain is generated by the Ethereum platform \cite{Ethereum}; Ripple is a real-time gross settlement system (RTGS), currency exchange and remittance network Ripple \cite{Ripple}, and IOTA is focused on providing secure communications and payments between agents on the Internet of Things \cite{Iota}. 

The emergence of a self-organised market of virtual currencies and/or assets whose value is generated primarily by social consensus~\cite{Baronchelli172189} has naturally attracted interest from the scientific community~\cite{dwyer2015economics,bohme2015bitcoin,casey2015bitcoin,trimborn2016crix,iwamura2014bitcoin,cusumano2014bitcoin,
wu2018classification,lamarche2018coexistence,krafft2018experimental,ali2014economics,
rogojanu2014issue,white2015market,ceruleo2014bitcoin,
sayed2018impact,javarone2018bitcoin,sovbetov2018factors,
parino2018analysis,beguvsic2018scaling}. Recent results have shown that the long-term properties of the cryptocurrency marked have remained stable between 2013 and 2017 and are compatible with a scenario in which investors simply sample the market and allocate their money according to the cryptocurrency's market shares~\cite{ElBahrawy170623}. While this is true on average, various studies have focused on the analysis and forecasting of price fluctuations, using mostly traditional approaches for financial markets analysis and prediction~\cite{ciaian2016economics,guo2018predicting,gajardo2018does,gandal2016can,
elendner2016cross}.

The success of machine learning techniques for stock markets prediction~\cite{enke2005use, huang2005forecasting, ou2009prediction, gavrilov2000mining, kannan2010financial, sheta2015comparison, chang2009ensemble}, suggests that these methods could be effective also in predicting cryptocurrencies prices.  However, the application of machine learning algorithms to the cryptocurrency market has been limited so far to the analysis of Bitcoin prices, using random forests~\cite{madan2015automated}, Bayesian neural network~\cite{jang2018empirical}, long short-term memory neural network~\cite{mcnally2016predicting} and other algorithms~\cite{hegazycomparitive,guo2018predicting}. These studies were able to anticipate, to different degrees, the price fluctuations of Bitcoin, and revealed that best results were achieved by neural network based algorithms. Deep reinforcement learning was showed to beat the uniform buy and hold strategy~\cite{shilling1992market} in predicting the prices of $12$ cryptocurrencies over one year period ~\cite{jiang2016cryptocurrency}. 

Other attempts to use machine learning to predict the prices of cryptocurrencies other than Bitcoin come from non-academic sources~\cite{traderobot,CryptoBot,CryptoCurrencyTrader,btctrading,bitpredict,
bitcoinbubbleburst}. Most of these analyses focused on a limited number of currencies and did not provide benchmark comparisons for their results.

Here, we test the performance of three models in predicting daily cryptocurrency price for 1,681 currencies. Two of the models are based on gradient boosting decision trees~\cite{friedman2002stochastic} and one is based on long short-term memory (LSTM) recurrent neural networks~\cite{hochreiter1997long}. In all cases, we build investment portfolios based on the predictions and we compare their performance in terms of return on investment. We find that all of the three models perform better than a baseline `simple moving average' model \cite{brock1992simple,kilgallen2012testing,lebaron2000stability,ellis2005smarter} where a currency's price is predicted as the average price across the preceding days,  and that the method based on long short-term memory recurrent neural networks systematically yields the best return on investment. 

The article is structured as follows: In section \nameref{sec:Methods} we describe the data (see \nameref{subsec:data}), the metrics characterizing cryptocurrencies that are used along the paper (see \nameref{sec:Metrics}), the forecasting algorithms (see \nameref{sec:algorithms}), and the evaluation metrics (see \nameref{sec:evaluation}). In section \nameref{sec:Results}, we present and compare the results obtained with the three forecasting algorithms and the baseline method. In section \nameref{sec:Discussion}, we conclude and discuss results.

\section*{Materials and Methods}\label{sec:Methods}
\subsection*{Data description and pre-processing}\label{subsec:data}
\begin{figure}[htb]
\centering
\includegraphics[width=0.8\textwidth]{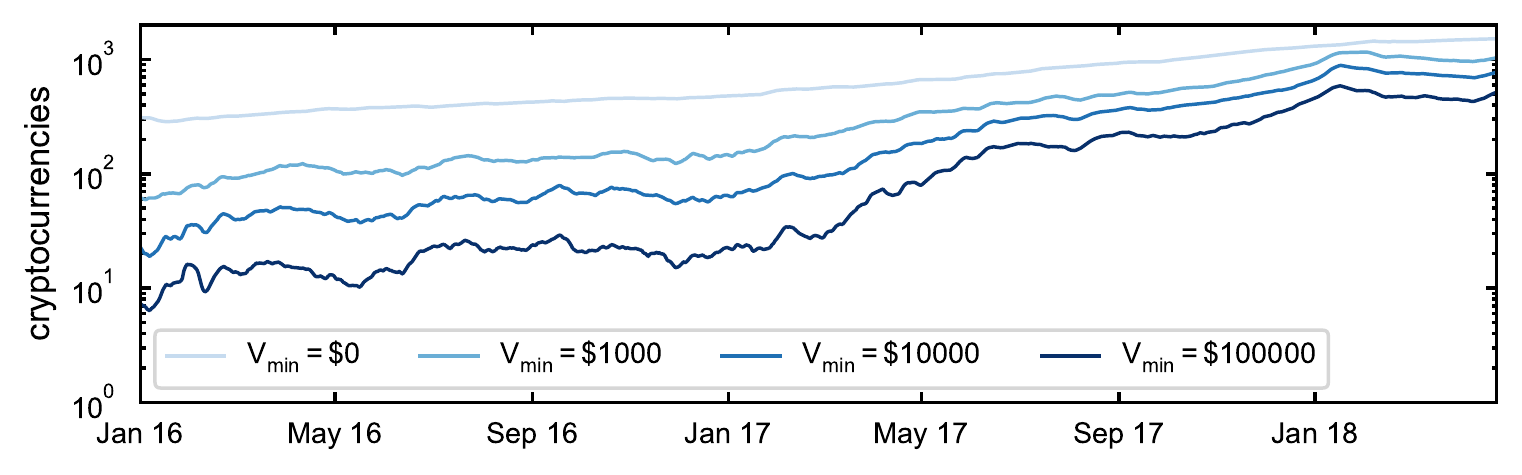}
\caption{\textbf{Number of cryptocurrencies.} The cryptocurrencies with volume higher than $V_{min}$ as a function of time, for different values of  $V_{min}$. For visualization purposes, curves are averaged over a rolling window of $10$ days.}
\label{fig:min_volume}
\end{figure}

Cryptocurrency data was extracted from the website Coin Market Cap~\cite{coincap}, collecting daily data from $300$ exchange markets platforms starting in the period between November 11, 2015 and April 24, 2018. The dataset contains the daily price in U.S. dollars, the market capitalisation and the trading volume of $1,681$ cryptocurrencies, where the market capitalization is the product between price and circulating supply, and the volume is the number of coins exchanged in a day. The daily price is computed as the volume weighted average of all prices reported at each market.
\cref{fig:min_volume} shows the number of currencies with trading volume larger than $V_{min}$ over time, for different values of $V_{min}$. In the following sections, we consider that only currencies with daily trading volume higher than $10^5$ USD can be traded at any given day. 

The website lists cryptocurrencies traded on public exchange markets that have existed for more than 30 days and for which an API as well as a public URL showing the total mined supply are available. Information on the market capitalization of cryptocurrencies that are not traded in the 6 hours preceding the weekly release of data is not included on the website. Cryptocurrencies inactive for $7$ days are not included in the list released. These measures imply that some cryptocurrencies can disappear from the list to reappear later on. In this case, we consider the price to be the same as before disappearing. However, this choice does not affect results since only in 28 cases the currency has volume higher than $10^5$ USD right before disappearing (note that there are 124,328 entries in the dataset with volume larger than $10^5$ USD).

\subsection*{Metrics} 
\label{sec:Metrics}

Cryptocurrencies are characterised over time by several metrics, namely  
\begin{itemize}
\item{Price,}  The exchange rate, determined by supply and demand dynamics. 
\item{Market capitalization,}  The product of the circulating supply and the price. 
\item{Market share,} The market capitalization of a currency normalized by the total market capitalization.
\item{Rank,}  The \emph{rank} of currency based on its market capitalization. 
\item{Volume,}  Coins traded in the last 24 hours. 
\item{Age,}  Lifetime of the currency in days. 
\end{itemize}

The profitability of a currency $c$ over time can be quantified through the \emph{return on investment} (ROI), measuring the return of an investment made at day $t_i$ relative to the cost~\cite{friedlob1996understanding}. The index $i$ rolls across days and it is included between $0$ and $844$, with $t_0=$ January 1, 2016, and $t_{844}=$ April 24, 2018. Since we are interested in the short-term performance, we consider the return on investment after $1$ day defined as

\begin{equation}
ROI(c,t_i) = \frac{price(c,t_i)- price(c,t_i-1)}{price(c,t_i-1)} .
\label{eq:roi}
\end{equation}

\begin{figure}[htb]
\centering
\includegraphics[width=0.8\textwidth]{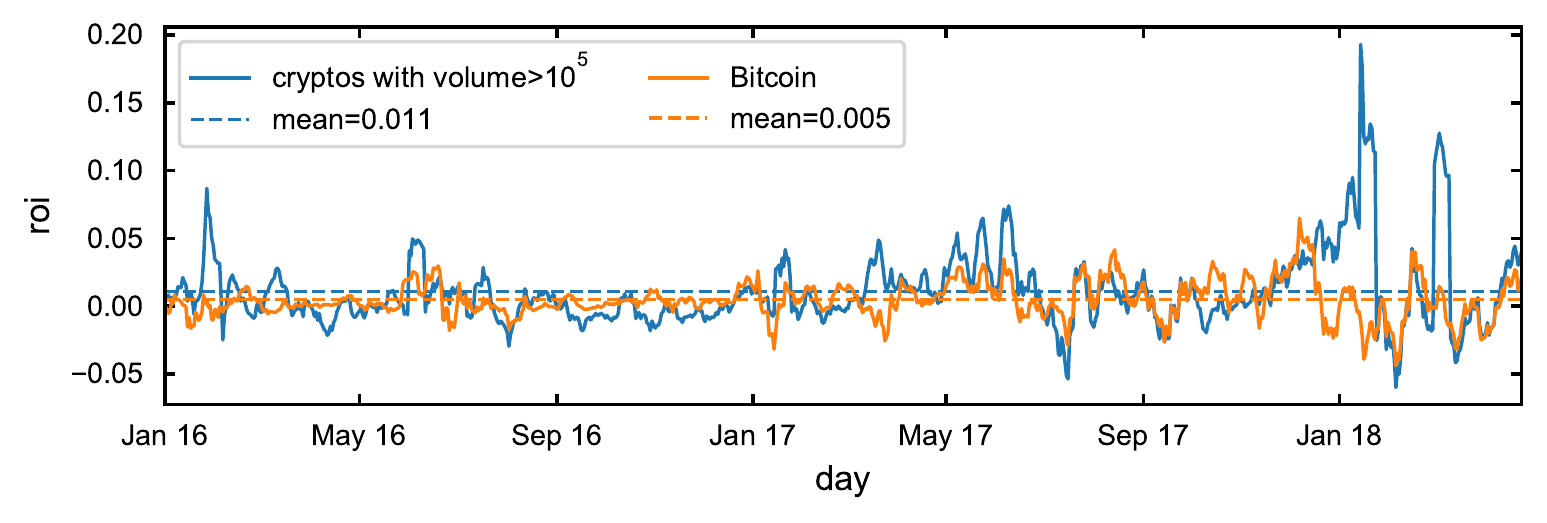}
\caption{\textbf{Return on investment over time.} The daily return on investment for Bitcoin (orange line) and the average for currencies with volume larger than $V_{min} = 10^5$ USD (blue line). Their average value across time (dashed lines) is larger than $0$. For visualization purposes, curves are averaged over a rolling window of $10$ days. }
\label{fig:roi}
\end{figure}

In \cref{fig:roi}, we show the evolution of the $ROI$ over time for Bitcoin (orange line) and on average for currencies whose volume is larger than $V_{min} = 10^5$ USD at $t_i-1$ (blue line). In both case, the average return on investment over the period considered is larger than $0$, reflecting the overall growth of the market. 

\subsection*{Forecasting algorithms} 
\label{sec:algorithms}

We test and compare three supervised methods for short-term price forecasting. The first two methods rely on XGboost~\cite{chen2016xgboost}, an open-source scalable machine learning system for tree boosting used in a number of winning Kaggle solutions (17/29 in 2015) \cite{Kaggle}. The third method is based on the long short-term memory (LSTM) algorithm for recurrent neural networks~\cite{hochreiter1997long} that have demonstrated to achieve state-of-the-art results in time-series forecasting \cite{lipton2015critical}. 

\begin{figure}[htb]
\centering
\includegraphics[width=0.8\textwidth]{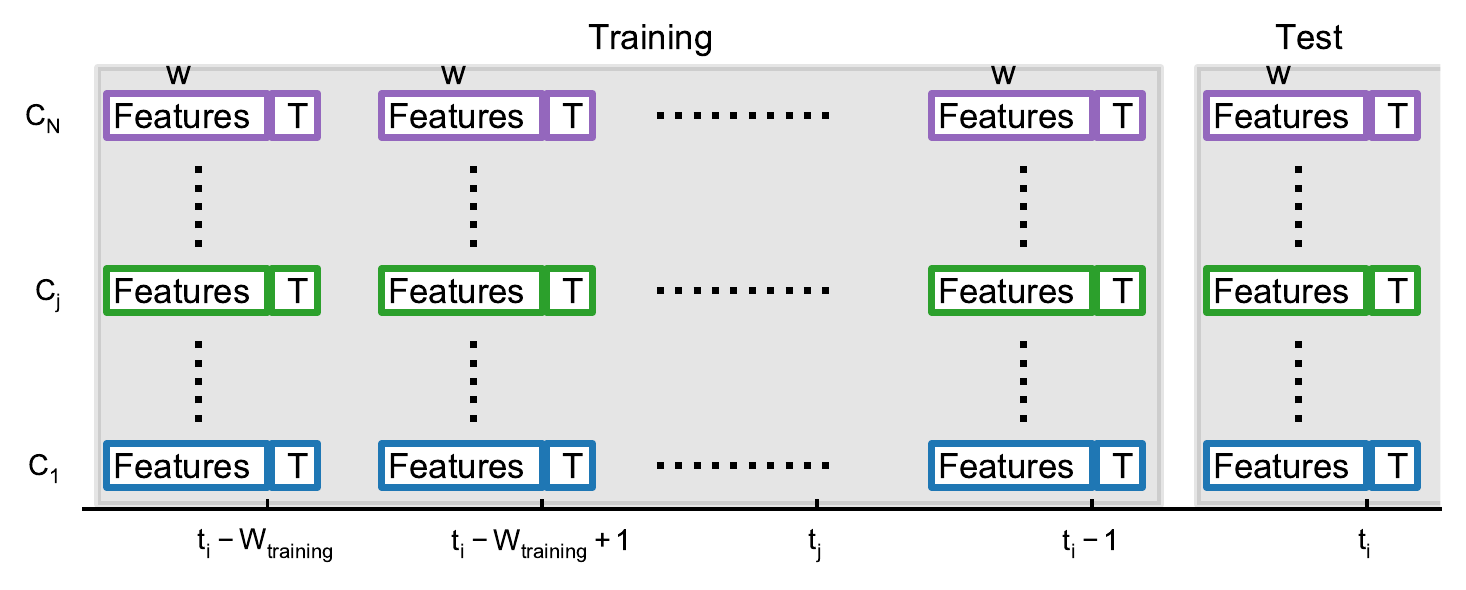}
\caption{\textbf{Schematic description of Method 1.} The training set is composed of features and target (T) pairs, where features are various characteristics of a currency $c_i$, computed across the $w$ days preceding time $t_j$ and the target $T$ is the price of $c_i$ at $t_j$. The features-target pairs are computed for all currencies $c_i$ and all values of $t_j$ included between $t_i-W_{training}$ and $t_i-1$. The test set includes features-target pairs for all currencies with trading volume larger than $10^5$ USD at $t_i$, where the target is the price at time $t_i$ and features are computed in the $w$ days preceding $t_i$. }
\label{schema}
\end{figure}

\textbf{Method 1:} The first method considers one single regression model to describe the change in price of all currencies (see \cref{schema}). The model is an ensemble of regression trees built by the XGboost algorithm. The features of the model are characteristics of a currency between time $t_j-w$ and $t_j-1$ and the target is the ROI of the currency at time $t_j$, where $w$ is a parameter to be determined. The characteristics considered for each currency are: price, market capitalization, market share, rank, volume and ROI (see equation \cref{eq:roi}). The features for the regression are built across the window between $t_j - w$ and $t_j-1$ included (see \cref{schema}). Specifically, we consider the average, the standard deviation, the median, the last value and the trend (e.g. the difference between last and first value) of the properties listed above. 
In the training phase, we include all currencies with volume larger than $10^5$ USD, and $t_j$ between $t_i-W_{training}$ and $t_i$. In general, larger training windows do not necessarily lead to better results (see results section), because the market evolves across time. In the prediction phase, we test on the set of existing currencies at day $t_i$. This procedure is repeated for values of $t_i$ included between January 1, 2016 and April 24, 2018.

\begin{figure}[!h]
\centering
\includegraphics[width=0.8\textwidth]{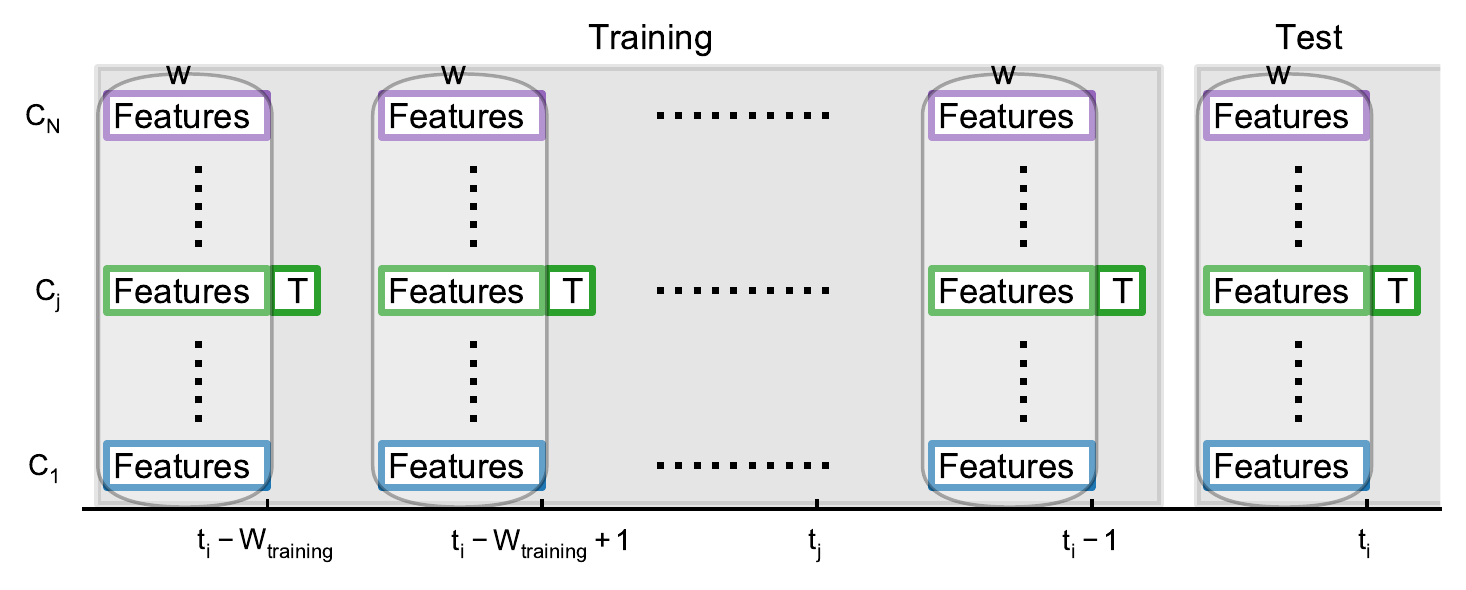}
\caption{\textbf{Schematic description of Method 2.} The training set is composed of features and target (T) pairs, where features are various characteristics of all currencies, computed across the $w$ days preceding time $t_j$ and the target $T$ is the price of $c_i$ at $t_j$. The features-target pairs include a single currency $c_i$, for all values of $t_j$ included between $t_i-W_{training}$ and $t_i-1$. The test set contains a single features-target pair: the characteristics of all currencies, computed across the $w$ days preceding time $t_i$ and the price of $c_i$ at $t_i$. }\label{schema2}
\end{figure}

\textbf{Method 2:} Also the second method relies on XGboost, but now the algorithm is used to build a different regression model for each currency $c_i$ (see \cref{schema2}). The features of the model for currency $c_i$ are the characteristics of all the currencies in the dataset between $t_j-w$ and $t_j-1$ included and the target is the ROI of $c_i$ at day $t_j$(i.e., now the algorithm learns to predict the price of the currency $i$ based on the features of all the currencies in the system between $t_j-w$ and $t_j-1$). The features of the model are the same used in Method 1 (e.g. the average, standard, deviation, median, last value, difference between last and first value of the following quantities: price, market capitalisation, market share, rank, volume and ROI) across a window of length $w$. The model for currency $c_i$ is trained with pairs features target between times $t_i-W_{training}$ and $t_i-1$. The prediction set include only one pair: the features (computed between $t_i-w$ and $t_i-1$) and the target (computed at $t_i$) of currency $c_i$.

\textbf{Method 3: }The third method is based on Long Short Term Memory networks, a special kind of Recurrent Neural Networks, capable of learning long-term dependencies. As for Method 2, we build a different model for each currency.  Each model predicts the ROI of a given currency at day $t_i$ based on the values of the ROI of the same currency between days $t_i-w$ and $t_i-1$ included. 

\textbf{Baseline method:} As baseline method, we adopt the simple moving average strategy (SMA) widely tested and used as a null model in stock market prediction~\cite{brock1992simple,kilgallen2012testing,lebaron2000stability,ellis2005smarter}. It estimates the price of a currency at day $t_i$ as the average price of the same currency between $t_i-w$ and $t_i-1$ included.

\subsection*{Evaluation}
\label{sec:evaluation}

We compare the performance of various investment portfolios built based on the algorithms predictions. The investment portfolio is built at time $t_i-1$ by equally splitting an initial capital among the top $n$ currencies predicted with positive return. Hence, the total return at time $t_i$ is:
\[
R(t_i) =\frac{1}{n} \sum_{c=1}^{n} ROI (c,t_i).
\]
 
\noindent The portfolios performance is evaluated by computing the Sharpe ratio and the geometric mean return. The Sharpe ratio is defined as: 
\[
S(t_i) =\frac{\overline{R}}{s_{R}},
\]
where $\overline{R}$ is the average return on investment obtained between times $0$ and $t_i$, and $s_{R}$, the corresponding standard deviation. 

\noindent The geometric mean return is defined as:
\[
G(t_i) =\sqrt[T]{\prod_{t_j=1}^{t_i}{1+R(t_j)}},
\]

\noindent where $t_i$ corresponds to the total number of days considered. The cumulative return obtained at $t_i$ after investing and selling on the following day for the whole period is defined as $G(t_i)^2$.

The number of currencies $n$ to include in a portfolio is chosen at $t_i$ by optimising either the geometric mean $G(t_i-1)$ (geometric mean optimisation) or the Sharpe ratio $S(t_i-1)$ (Sharpe ratio optimisation) over the possible choices of $n$. The same approach is used to choose the parameters of Method 1 ($w$ and $W_{training}$), Method 2 ($w$ and $W_{training}$), and the baseline method ($w$).

\section*{Results}
\label{sec:Results}
We predict the price of the currencies at day $t_i$, for all $t_i$ included between Jan, 1st 2016 and Apr 24th, 2018.  The analysis considers all currencies whose age is larger than $50$ days since their first appearance and whose volume is larger than $\$ 100000$. To discount for the effect of the overall market movement (i.e., market growth, for most of the considered period), we consider cryptocurrencies prices expressed in Bitcoin. This implies that Bitcoin is excluded from our analysis.

\subsection*{Parameter setting}
First, we choose the parameters for each method. Parameters include the number of currencies $n$ to include the portfolio as well as the parameters specific to each method. In most cases, at each day $t_i$ we choose the parameters that maximise either the geometric mean $G(t_i-1)$ (geometric mean optimisation) or the Sharpe ratio $S(t_i-1)$ (Sharpe ratio optimisation) computed between times $0$ and $t_i$.

\textbf{Baseline strategy:} We test the performance of the baseline strategy for choices of window $w \geq 2$ (the minimal requirement for the $ROI$ to be different from $0$) and $w<30$. We find that the value of $w$ mazimising the geometric mean return (see Appendix \cref{baseline_btc}-A) and the Sharpe Ratio (see Appendix \cref{baseline_btc}-D) fluctuates especially before November $2016$ and has median value $4$ in both cases. The number of currencies included in the portfolio oscillates between $1$ and $11$ with median at $3$, both for the Sharpe Ratio (see \cref{baseline_btc}-B) and the geometric mean return (see \cref{baseline_btc}-E) optimisation.

\textbf{Method 1:} We explore values of the window $w$ in $\{3,5,7,10\}$ days and the training period $W_{training}$ in $\{5,10,20\}$ days (see Appendix \cref{method1}). We find that the median value of the selected window $w$ across time is $7$ for both the Sharpe ratio and the geometric mean optimisation. The median value of $W_{training}$ is $5$ under geometric mean optimisation and $10$ under Sharpe ratio optimisation. The number of currencies included in the portfolio oscillates between $1$ and $43$ with median at $15$ for the Sharpe Ratio (see Appendix \cref{method1}-A) and $9$ for the geometric mean return (see Appendix \cref{method1}-C) optimisations.

\textbf{Method 2:} We explore values of the window $w$ in $\{3,5,7,10\}$ days and the training period $W_{training}$ in $\{5,10,20\}$ days (see Appendix \cref{method2}). The median value of the selected window $w$ across time is $3$ for both the Sharpe ratio and the geometric mean optimisation. The median value of $W_{training}$ is 10 under geometric mean and Sharpe ratio optimisation. The number of currencies included has median at $17$ for the Sharpe Ratio and $7$ for the geometric mean optimisation  (see Appendix \cref{method2}-A and C).

\textbf{Method 3: }The LSTM has three parameters: The number of epochs, or complete passes through the dataset during the training phase; the number of neurons in the neural network, and the length of the window $w$. These parameters are chosen by optimising the price prediction of three currencies (Bitcoin, Ripple, and Ethereum) that have on average the largest market share across time (excluding Bitcoin Cash that is a fork of Bitcoin). Results (see SI,  \cref{parameters_optimisation_m3}) reveal that, in the range of parameters explored, the best results are achieved for $w=50$. Results are not particularly affected by the choice of the number of neurones nor the number of epochs. We choose $1$ neuron and 1000 epochs since the larger these two parameters, the larger the computational time.
The number of currencies to include in the portfolio is optimised over time by mazimising the geometric mean return (see Appendix \cref{method3}-A) and the Sharpe ratio (see Appendix \cref{method3}-B). In both cases the median number of currencies included is $1$.

\subsection*{Cumulative return}
\begin{figure}[htb]
\centering
\includegraphics[width=0.9\textwidth]{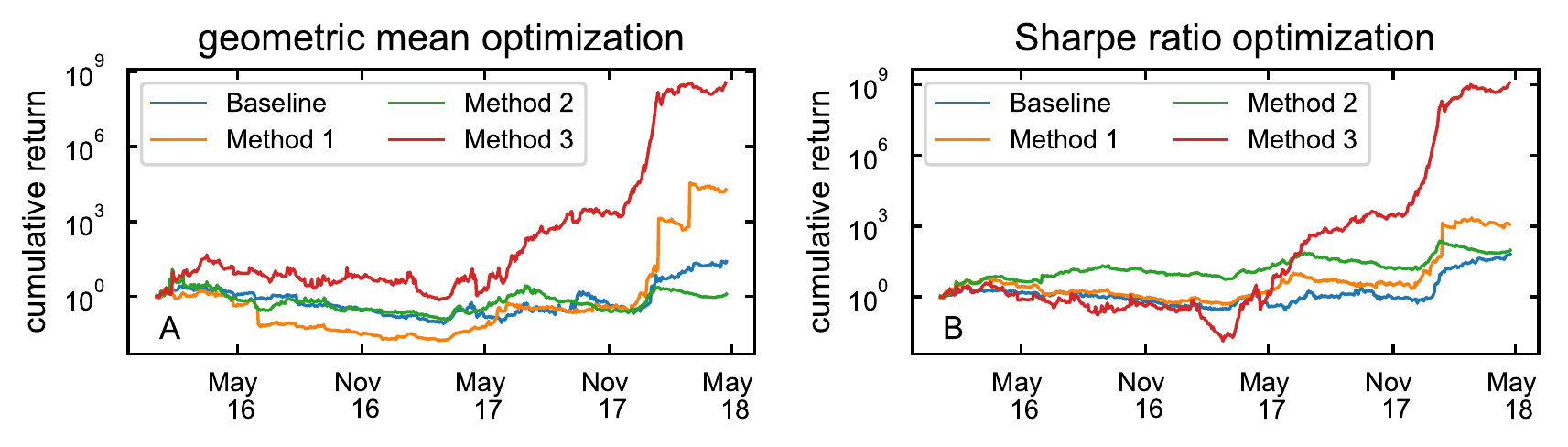}
\caption{\textbf{Cumulative returns.} The cumulative returns obtained under the Sharpe Ratio optimisation (A) and the geometric mean optimisation (B) for the baseline (blue line), Method 1 (orange line), Method 2 (green line) and Method 3 (red line). Analyses are performed considering prices in BTC.}
\label{cumulative}
\end{figure}

In \cref{cumulative}, we show the cumulative return obtained using the 4 methods. The cumulative returns achieved on April,24 under the Sharpe Ratio optimisation are $\sim 65$ BTC (Baseline), $\sim 1.1 \cdot 10^3$ BTC (Method 1),  $\sim 95$ BTC (Method 2),  $\sim 1.2 \cdot 10^9$ BTC (Method 3). Under geometric mean optimisation we obtain $\sim 25$ BTC (Baseline), $\sim 19 \cdot 10^3$ BTC (Method 1),  $\sim 1.25$ BTC (Method 2), $\sim 3.6 \cdot 10^8$ BTC (Method 3). The cumulative returns obtained in USD are higher (see \cref{cumulative_usd}). This is expected, since the Bitcoin price has increased during the period considered. While some of these figures appear exaggerated, it is worth noticing that (i) we run a theoretical exercise assuming that the availability of Bitcoin is not limited and (ii) under this assumption the upper bound to our strategy, corresponding to investing every day in the most performing currency results in a total cumulative return of $6 \cdot 10^{123}$ BTC (see Appendix \cref{future}). 
We consider also the more realistic scenario of investors paying a transaction fee when selling and buying currencies (see Appendix \cref{fees_appendix}). In most exchange markets, the fee is typically included between $0.1 \%$ and $0.5\%$ of the traded amount \cite{wikiFees}. For fees up to $0.2\%$, all the investment methods presented above lead, on average, to positive returns over the entire period (see \cref{fees_table}). The best performing method, Method 3, achieves positive gains also when fees up to $1\%$ are considered (see \cref{fees_table}). 

\begin{figure}[htb]
\centering
\includegraphics[width=0.7\textwidth]{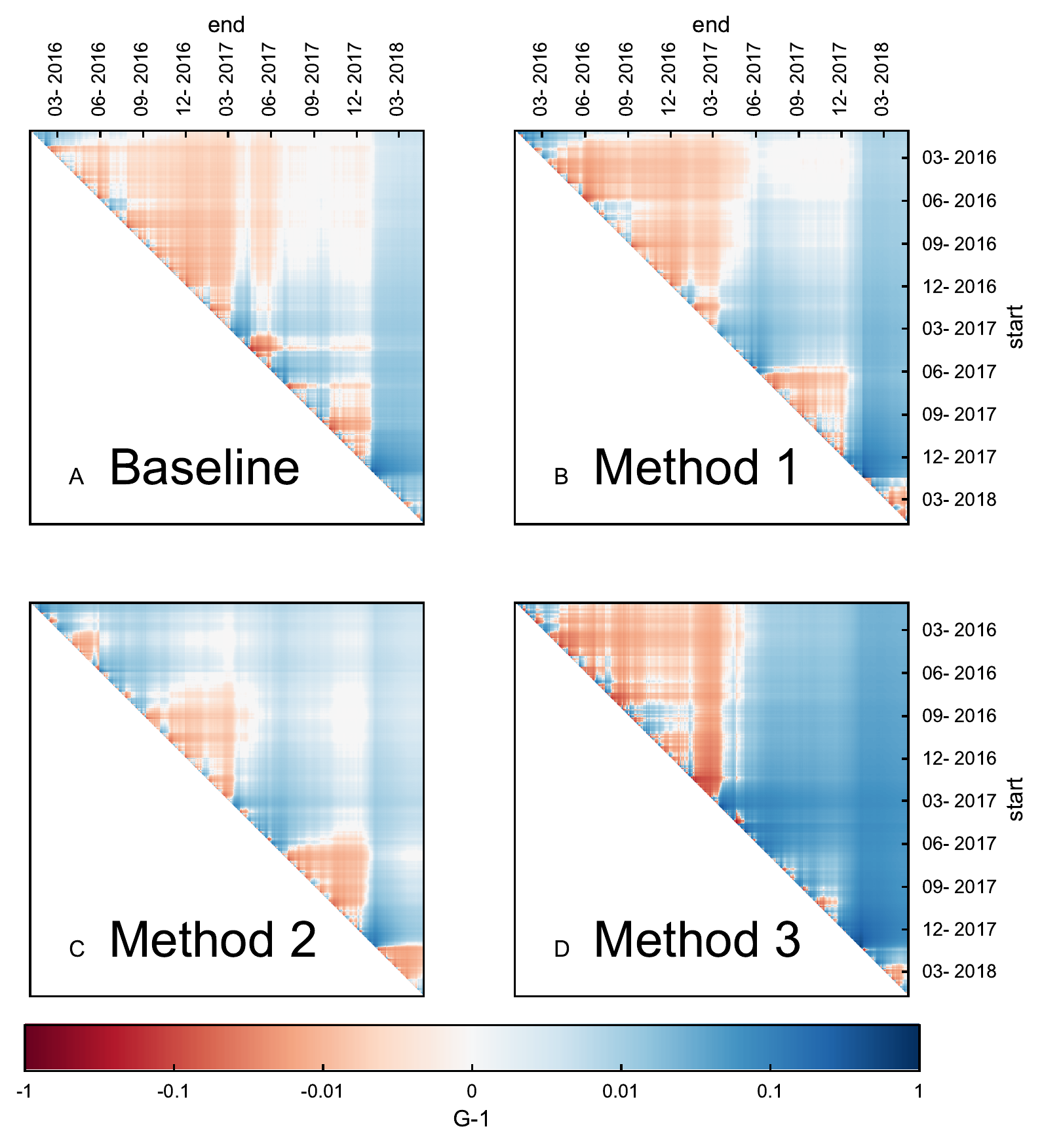}
\caption{\textbf{Geometric mean return obtained within different periods of time.} The geometric mean return computed between time "start" and "end" using the Sharpe ratio optimisation for the baseline (A), Method 1 (B), Method 2 (C) and Method 3 (D). Note that, for visualization purposes, the figure shows the translated geometric mean return G-1. Shades of red refers to negative returns and shades of blue to positive ones (see colour bar). }
\label{matrix_sharpe}
\end{figure}

The cumulative return in \cref{cumulative} is obtained by investing between January 1st, 2016 and April 24th, 2018. We investigate the overall performance of the various methods by looking at the geometric mean return obtained in different periods (see \cref{matrix_sharpe}). Results presented in \cref{matrix_sharpe} are obtained under Sharpe ratio optimisation for the baseline (\cref{matrix_sharpe}-A), Method 1 (\cref{matrix_sharpe}-B), Method 2 (\cref{matrix_sharpe}-C), and Method 3 (\cref{matrix_sharpe}-D). Note that, while in this case the investment can start after January 1st, 2016, we optimised the parameters by using data from that date on in all cases. Results are considerably better than those achieved using geometric mean return optimisation (see Appendix \cref{matrix_geom}). Finally, we observe that better performance is achieved when the algorithms consider prices in Bitcoin rather than USD (see \cref{table_USD}).

\subsection*{Feature importance}

\begin{figure}[!h]
\centering
\includegraphics[width=0.8\textwidth]{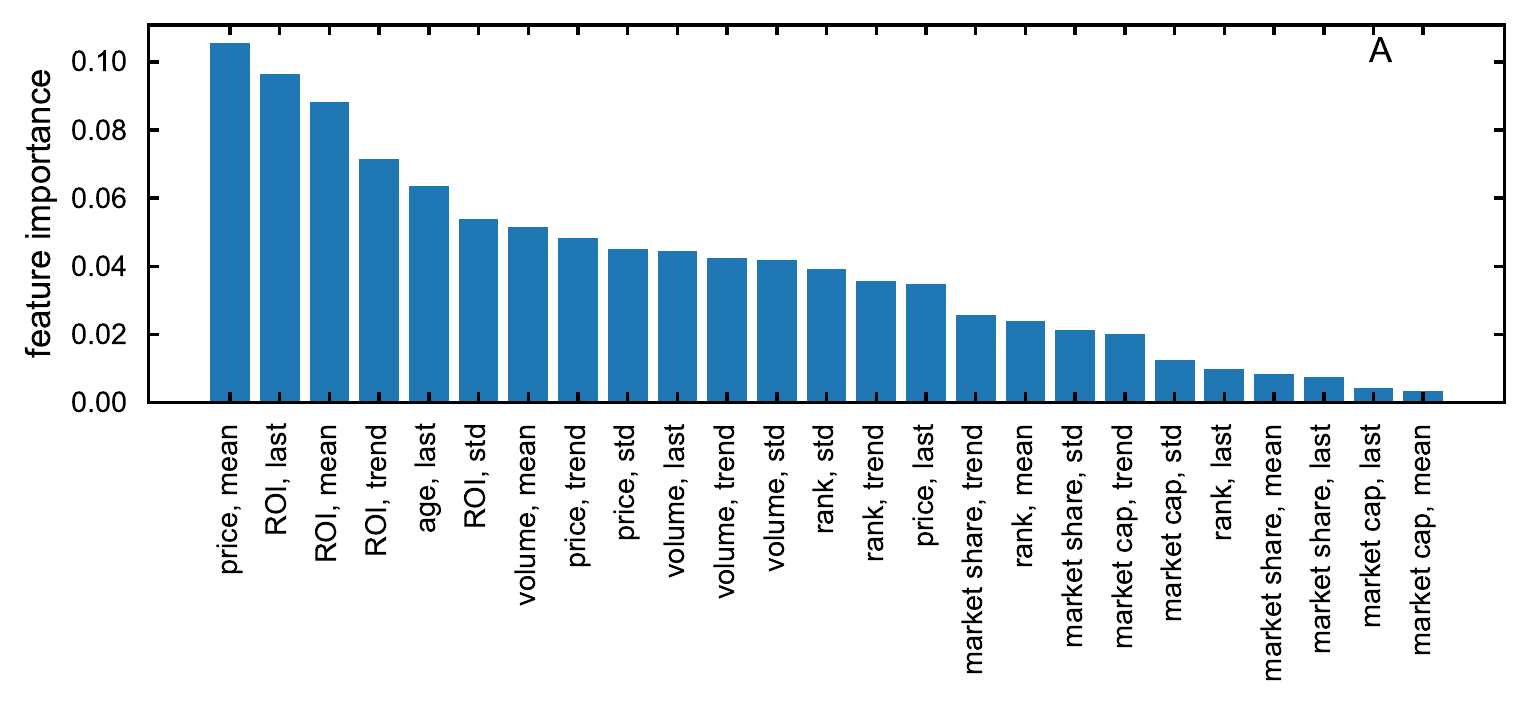}
\includegraphics[width=0.8\textwidth]{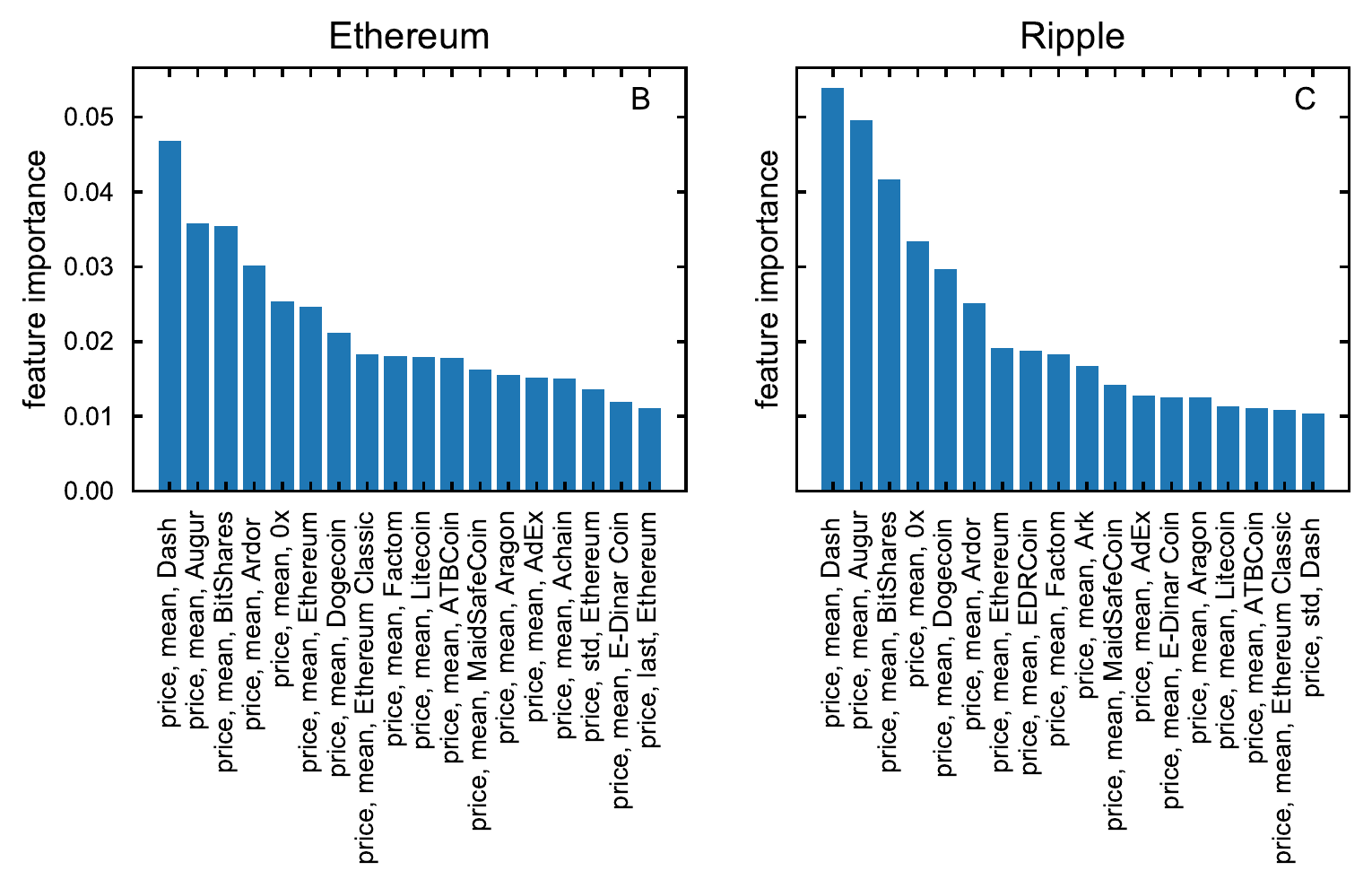}
\caption{\textbf{Feature importance for Methods 1 and 2.} (A) The average importance of each feature for the XGBoost regression model of Method 1.  Results are shown for $w=7$ and $W_{training} = 10$. (B,C) Examples of average feature importance for the XGBoost regression model developed in Method 2. Results are shown for $w=3$, $W_{training} = 10$, for Ethereum (B) and Ripple (C). For visualization purposes, we show only the top features.}
\label{feature_importance}
\end{figure}

In \cref{feature_importance}, we illustrate the relative importance of the various features in Method 1 and Method 2. For Method 1, we show the average feature importance; For Method 2, we show the average feature importances for two sample currencies: Ethereum and Ripple.

\subsection*{Portfolio composition}

The 10 most selected currencies under Sharpe Ratio optimisation are the following:

\textbf{Baseline: }{
Factom (91 days),
E-Dinar Coin (89 days),
Ripple (76 days),
Ethereum (71 days),
Steem (70 days),
Lisk (70 days),
MaidSafeCoin (69 days),
Monero (58 days),
BitShares (55 days),
EDRCoin (52 days).}

\textbf{Method 1: }{Ethereum (154 days), Dash (128 days),
 Monero (111 days),
 Factom (104 days),
 Ripple (94 days),
 Litecoin (93 days),
 Dogecoin (92 days),
 Maid Safe Coin (86 days),
 BitShares (73 days),
 Tether (59 days)}

\textbf{Method 2: }{Ethereum (63 days),
Monero (61 days),
Factom (51 days),
Ripple (42 days),
Dash (40 days),
Maid Safe Coin (40 days),
Siacoin (30 days),
NEM (26 days),
NXT (26 days),
Steem (23 days).}

\textbf{Method 3: }{Factom (48 days), Monero (46 days), Ethereum (39 days),  Lisk (36 days), Maid Safe Coin (32 days), E-Dinar Coin (32 days),  BitShares (26 days), B3 Coin (26 days), Dash (25 days), Cryptonite (22 days).}

\section*{Conclusion}
\label{sec:Discussion}
We tested the performance of three forecasting models on daily cryptocurrency prices for $1,681$ currencies. Two of them (Method 1 and Method 2) were based on gradient boosting decision trees and one is based on long short-term memory recurrent neural networks (Method 3). In Method 1, the same model was used to predict the return on investment of all currencies; in Method 2, we built a different model for each currency, that uses information on the behaviour of the whole market to make a prediction on that single currency; in Method 3, we used a different model for each currency, where the prediction is based on previous prices of the currency. 

We built investment portfolios based on the predictions of the different method and compared their performance with that of a baseline represented by the well known simple moving average strategy. The parameters of each model were optimised for all but Method 3 on a daily basis, based on the outcome of each parameters choice in previous times. We used two evaluation metrics used for parameter optimisation: The geometric mean return and the Sharpe ratio. To discount the effect of the overall market growth, cryptocurrencies prices were expressed in Bitcoin. All strategies, produced profit (expressed in Bitcoin) over the entire considered period and for a large set of shorter trading periods (different combinations of start and end dates for the trading activity), also when transaction fees up to $0.2\%$ are considered. 

The three methods performed better than the baseline strategy when the investment strategy was ran over the whole period considered. The optimisation of parameters based on the Sharpe ratio achieved larger returns. Methods based on gradient boosting decision trees (Method 1 and 2) worked best when predictions were based on short-term windows of 5/10 days, suggesting they exploit well mostly short-term dependencies. Instead, LSTM recurrent neural networks worked best when predictions were based on $\sim 50$ days of data, since they are able to capture also long-term dependencies and are very stable against price volatility. They allowed to make profit also if transaction fees up to $1\%$ are considered. Methods based on gradient boosting decision trees allow to better interpret results. We found that the prices and the returns of a currency in the last few days preceding the prediction were leading factors to anticipate its behaviour. Among the two methods based on random forests, the one considering a different model for each currency performed best (Method 2). Finally, it is worth noting that the three methods proposed perform better when predictions are based on prices in Bitcoin rather than prices in USD. This suggests that forecasting simultaneously the overall cryptocurrency market trend and the developments of individual currencies is more challenging than forecasting the latter alone.

It is important to stress that our study has limitations. First, we did not attempt to exploit the existence of different prices on different exchanges, the consideration of which could open the way to significantly higher returns on investment. Second, we ignored intra-day price fluctuations and considered an average daily price. Finally, and crucially, we run a theoretical test in which the available supply of Bitcoin is unlimited and none of our trades influence the market. Notwithstanding these simplifying assumptions, the methods we presented were systematically and consistently able to identify outperforming currencies. Extending the current analysis by considering these and other elements of the market is a direction for future work.

A different yet promising approach to the study cryptocurrencies consists in quantifying the impact of public opinion, as measured through social media traces, on the market behaviour, in the same spirit in which this was done for the stock market~\cite{moat2013quantifying}. While it was shown that social media traces can be also effective predictors of Bitcoin\cite{kondor2014inferring,kristoufek2015main,Garcia150288,kristoufek2013bitcoin,
urquhart2018causes,garcia2014digital,wang2017buzz} and other currencies \cite{li2018sentiment} price fluctuations, our knowledge of their effects on the whole cryptocurrency market remain limited and is an interesting direction for future work. 




\bibliographystyle{IEEEtran}
\bibliography{ArticleNew}

\clearpage
\beginsupplement
\section*{Appendix}

\subsection{Parameter optimisation}
\begin{figure}[htb]
\centering
\includegraphics[width=0.8\textwidth]{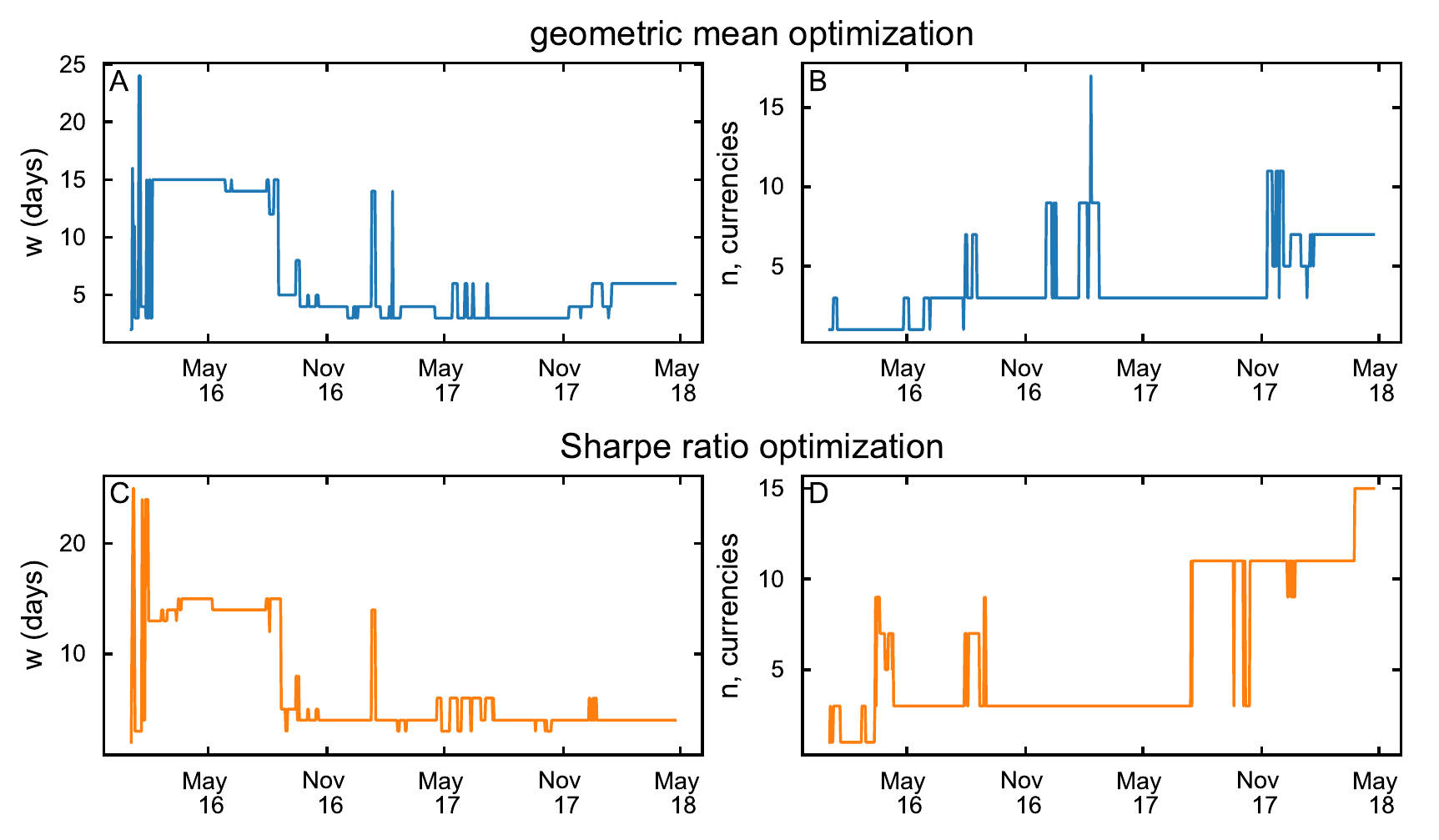}
\caption{\textbf{Baseline strategy: parameters optimisation.} The sliding window $w$ (A,C) and the number of currencies $n$ (B,D) chosen over time under the geometric mean (A,B) and the Sharpe Ratio optimisation (C,D). Analyses are performed considering prices in BTC.}
\label{baseline_btc}
\end{figure}

\begin{figure}[htb]
\centering
\includegraphics[width=0.8\textwidth]{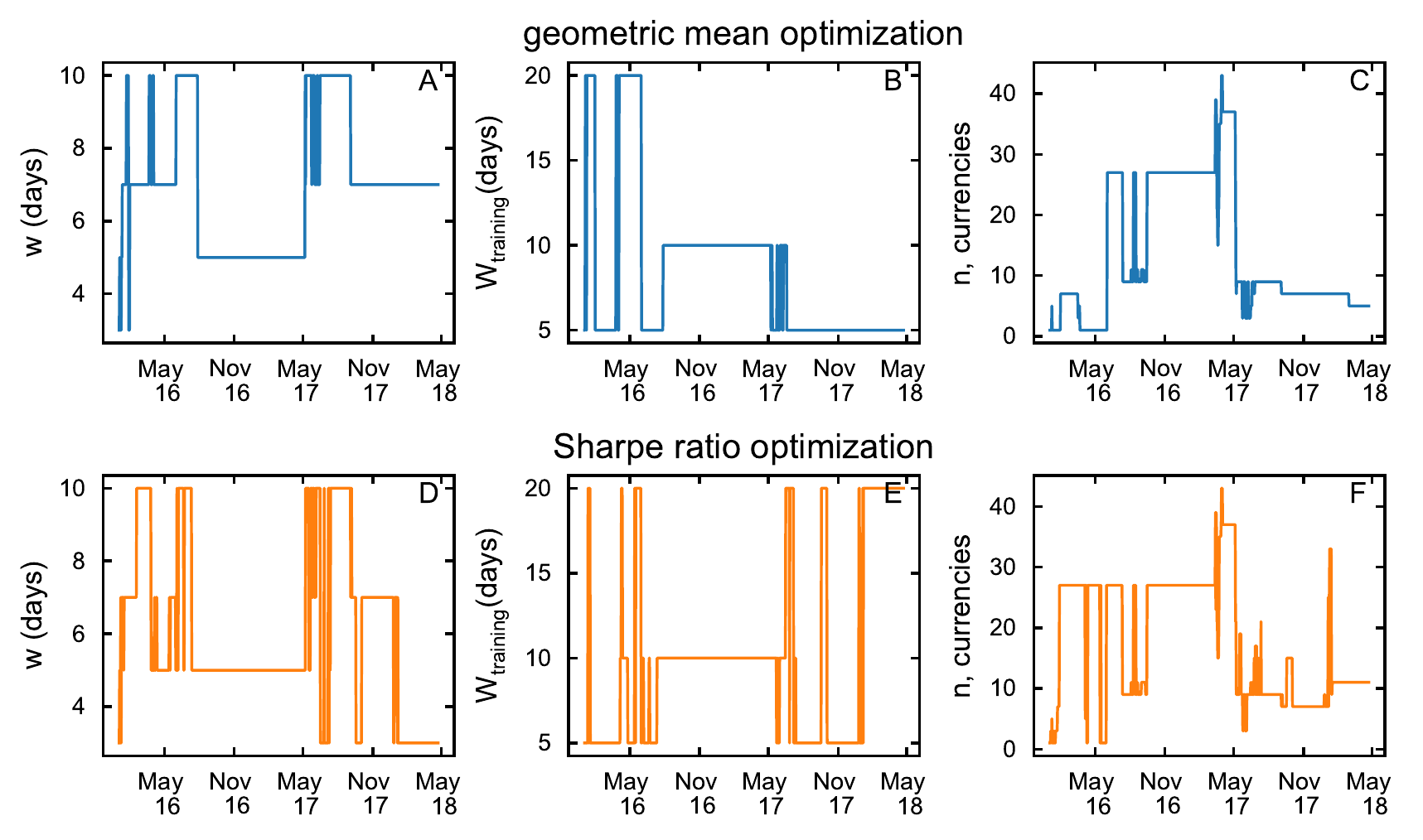}
\caption{\textbf{Method 1: Parameters optimisation.} The sliding window $w$ (A,D), the training window $W_{training}$ (B,E) and the number of currencies $n$ (C,F) chosen over time under the geometric mean (A,B,C) and the Sharpe Ratio optimisation (D,E,F). Analyses are performed considering prices in BTC.}
\label{method1}
\end{figure}

\begin{figure}[htb]
\centering
\includegraphics[width=0.8\textwidth]{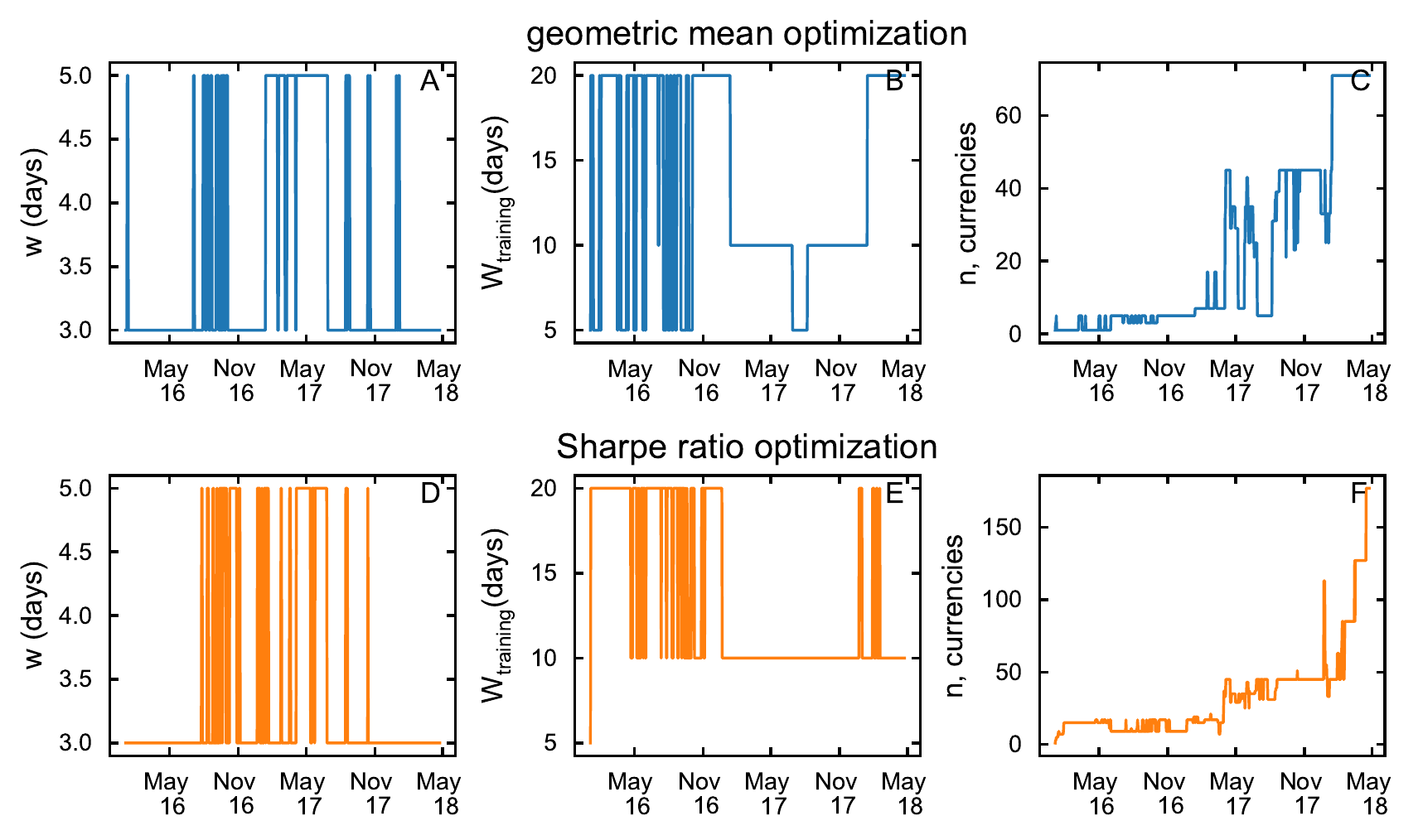}
\caption{\textbf{Method 2: Parameters optimisation.} The sliding window $w$ (A,D), the training window $W_{training}$ (B,E) and the number of currencies $n$ (C,F) chosen over time under the geometric mean (A,B,C) and the Sharpe Ratio optimisation (D,E,F). Analyses are performed considering prices in BTC.}
\label{method2}
\end{figure}

\begin{figure}[htb]
\centering
\includegraphics[width=0.7\textwidth]{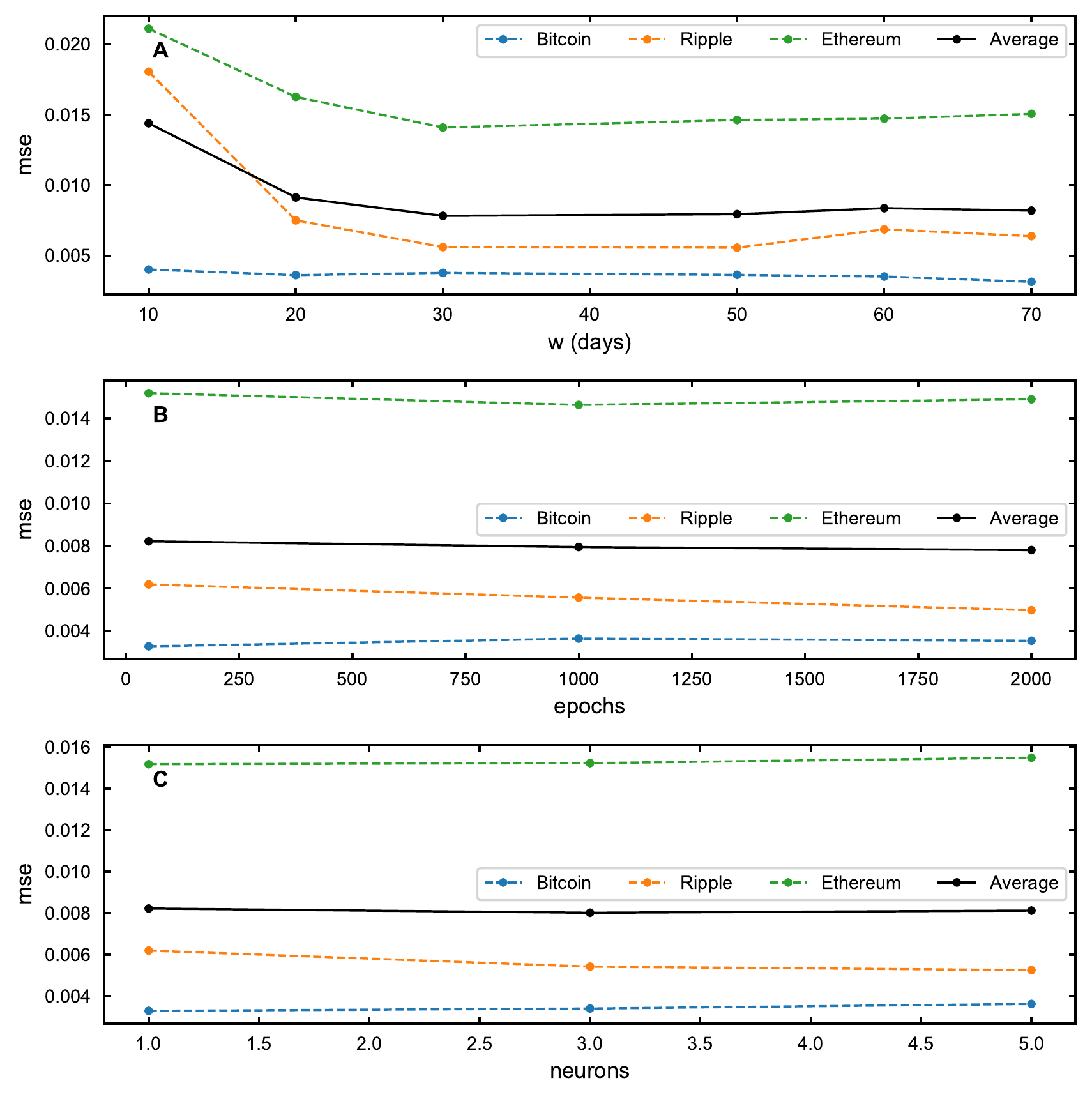}
\caption{\textbf{Method 3: Parameters optimisation.} The median squared error of the ROI as a function of the window size (A), the number of epochs (B) and the number of neurons (C). Results are shown considering prices in Bitcoin. }
\label{parameters_optimisation_m3}
\end{figure}

\begin{figure}[htb]
\centering
\includegraphics[width=0.8\textwidth]{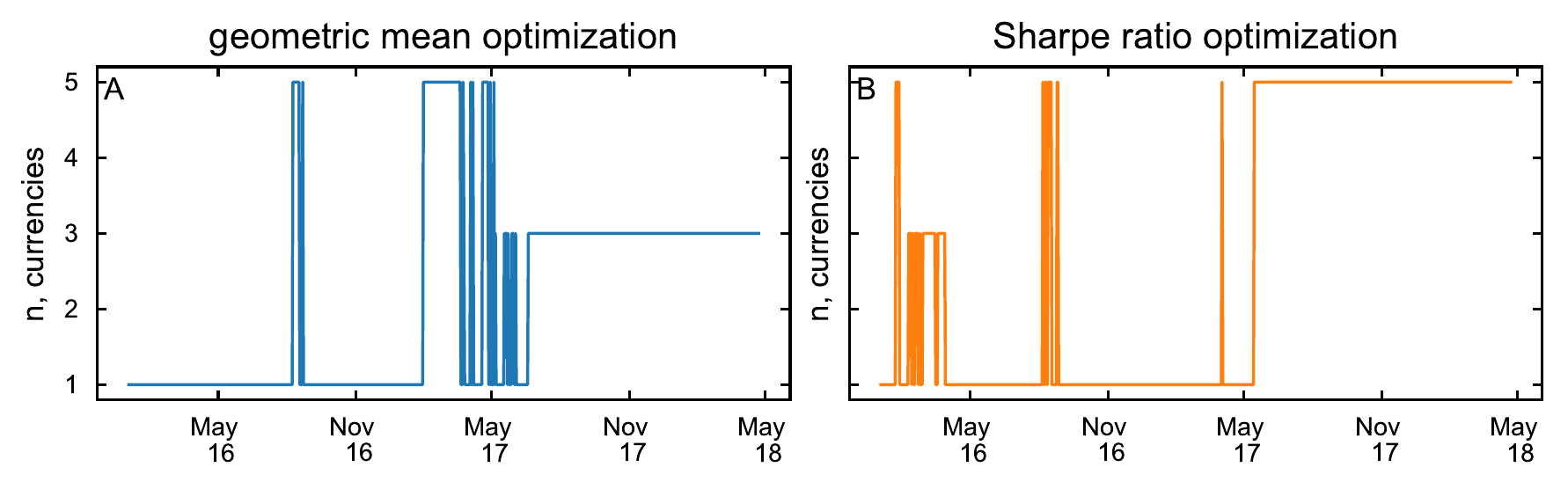}
\caption{\textbf{Method 3: Parameters optimisation. } The number of currencies $n$ chosen over time under the geometric mean (A) and the Sharpe Ratio optimisation (B). Analyses are performed considering prices in BTC.}
\label{method3}
\end{figure}

In \cref{baseline_btc}, we show the optimisation of the parameters $w$ (A,C) and $n$ (B,D) for the baseline strategy.
In \cref{method1}, we show the optimisation of the parameters $w$ (A,D), $W_{training}$ (B,E),  and $n$ (C,F) for Method 1.
In \cref{method2}, we show the optimisation of the parameters $w$ (A,D), $W_{training}$ (B,E),  and $n$ (C,F) for Method 2.
In \cref{parameters_optimisation_m3}, we show the median squared error obtained under different training window choices (A), number of epochs (B) and number of neurons (C), for Ethereum, Bitcoin and Ripple. 
In \cref{method3}, we show the optimisation of the parameter $n$ (C,F) for Method 3.

\subsection{Return under full knowledge of the market evolution.}
\begin{figure}[htb]
\centering
\includegraphics[scale = 0.8]{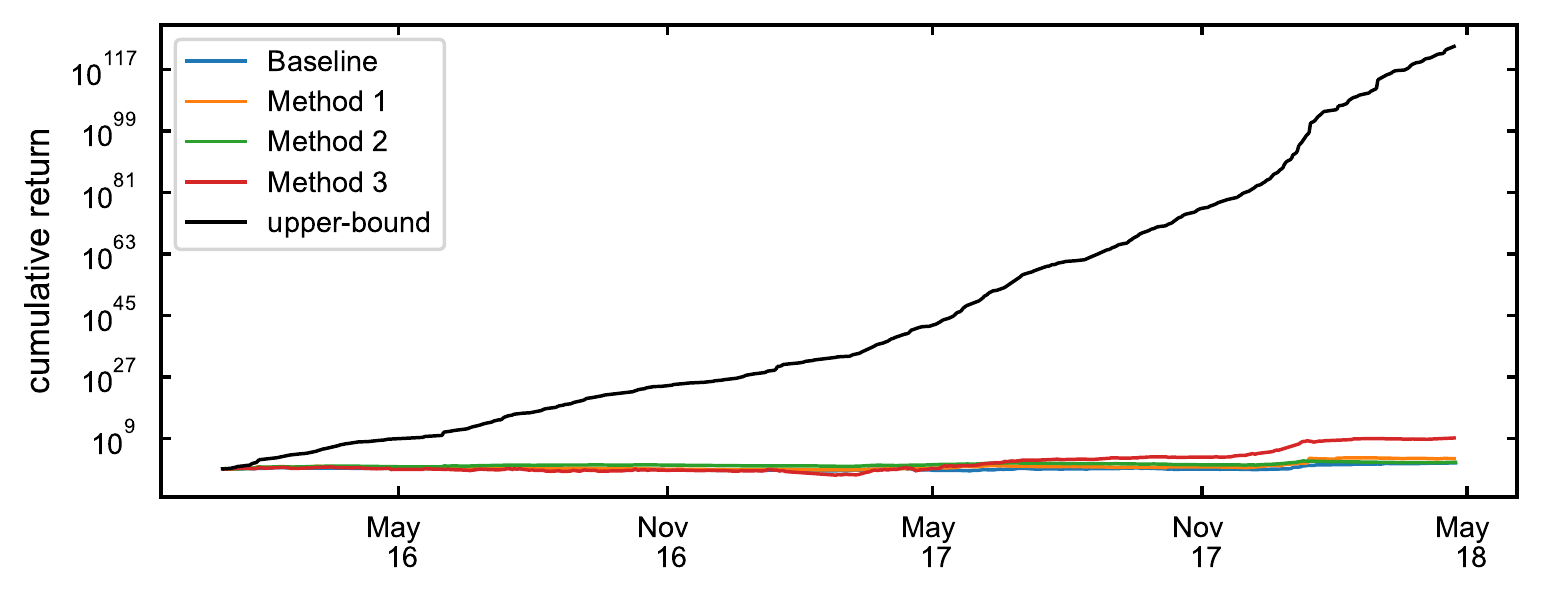}
\caption{\textbf{Upper-bound for the cumulative return.} The cumulative return obtained by investing every day in the currency with highest return on the following day (black line). The cumulative return obtained with the baseline (blue line), Method 1 (orange line), Method 2 (green line), and Method 3 (red line). Results are shown in Bitcoin. }
\label{future}
\end{figure}

In \cref{future}, we show the cumulative return obtained by investing every day in the top currency, supposing one knows the prices of currencies on the following day. 

\subsection{Return obtained paying transaction fees.}
\label{fees_appendix}
\begin{figure}[htb]
\centering
\includegraphics[width=0.7\textwidth]{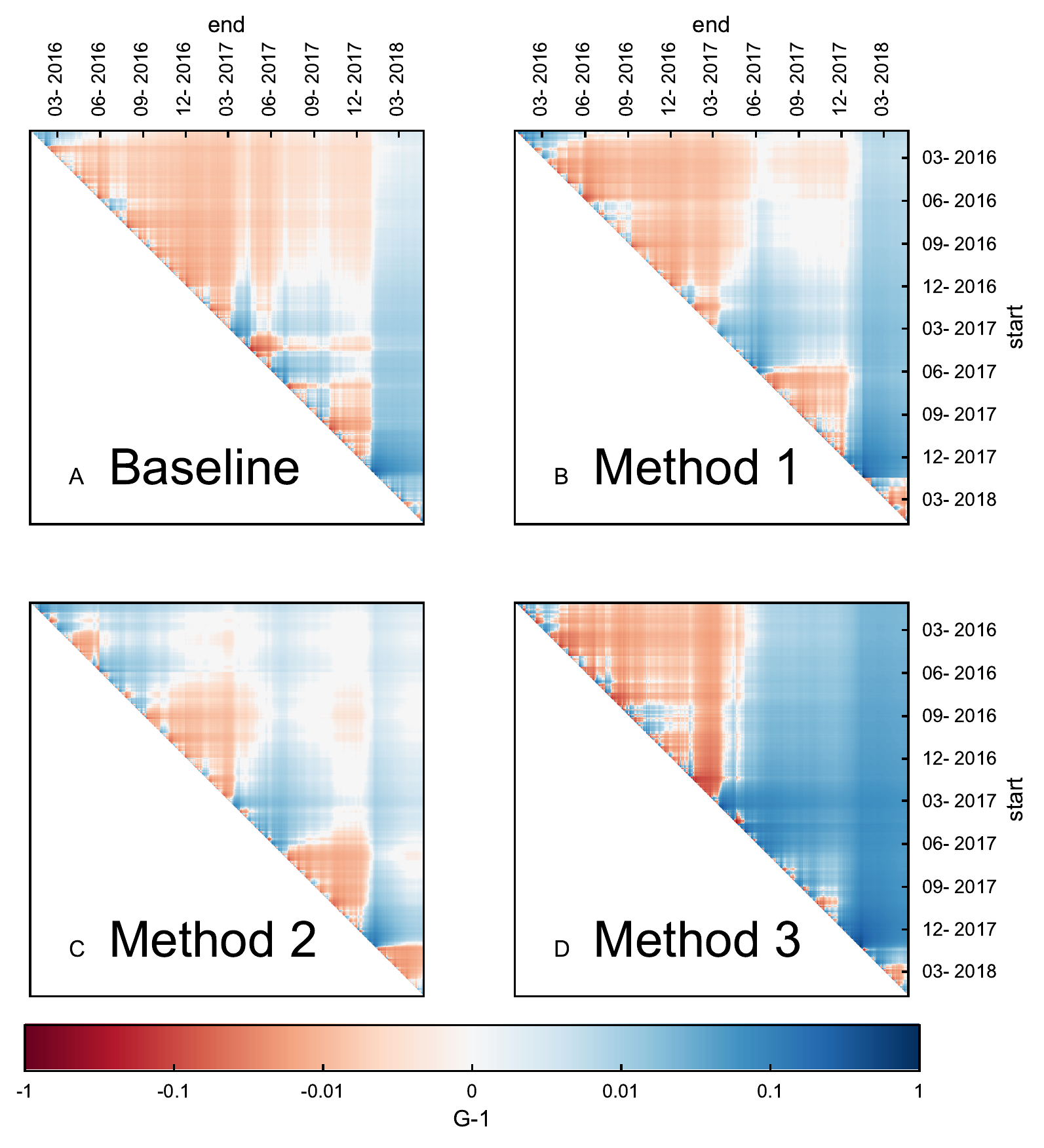}
\caption{\textbf{Daily geometric mean return obtained under transaction fees of $0.1\%$.} The geometric mean return computed between time "start" and "end" using the Sharpe ratio optimisation for the baseline (A), Method 1 (B), Method 2 (C) and Method 3 (D). Note that, for visualization purposes, the figure shows the translated geometric mean return G-1. Shades of red refers to negative returns and shades of blue to positive ones (see colour bar). }
\label{matrix_sharpe_fee_1}
\end{figure}

\begin{figure}[htb]
\centering
\includegraphics[width=0.7\textwidth]{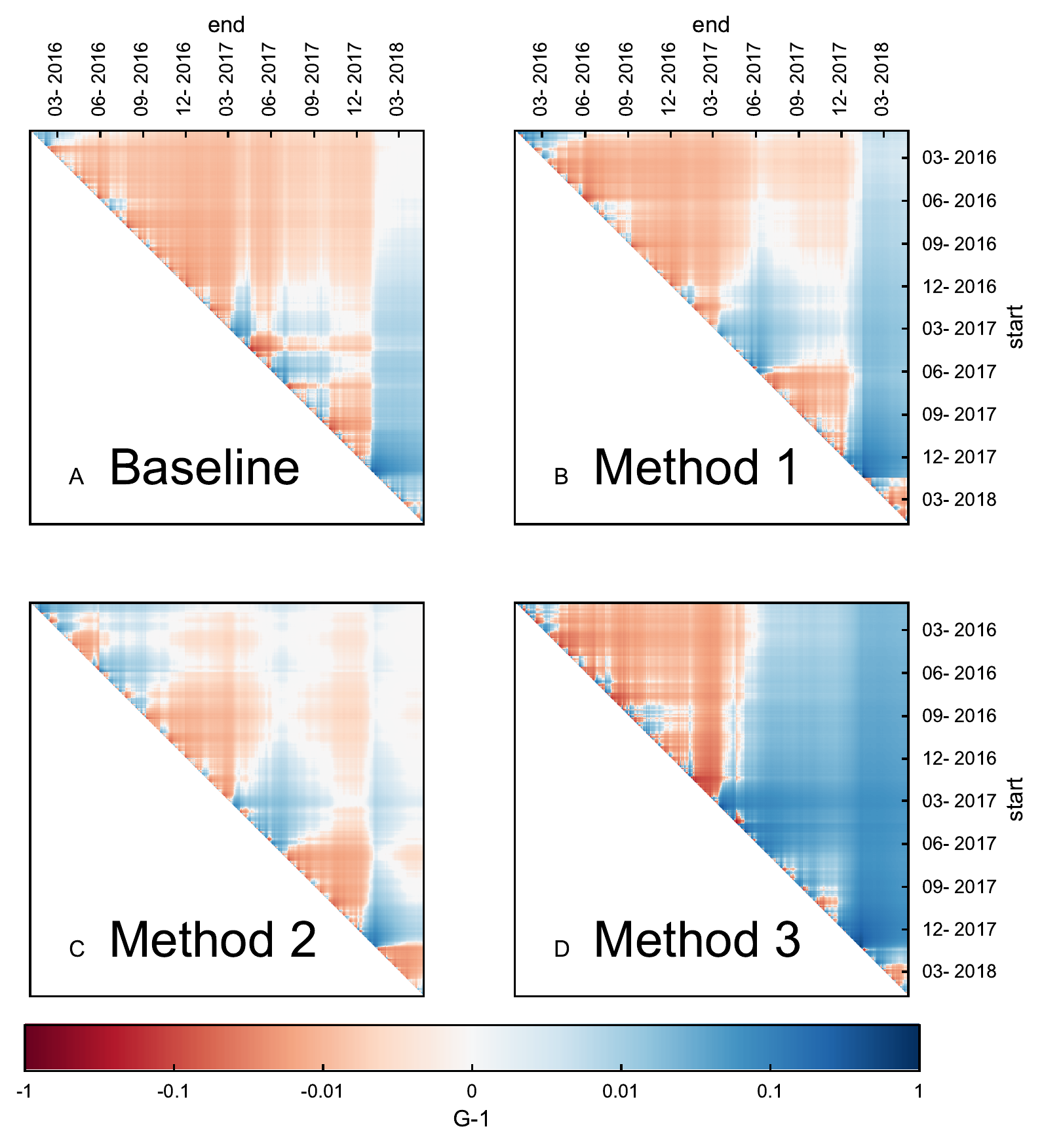}
\caption{\textbf{Daily geometric mean return obtained under transaction fees of $0.2\%$.} The geometric mean return computed between time "start" and "end" using the Sharpe ratio optimisation for the baseline (A), Method 1 (B), Method 2 (C) and Method 3 (D). Note that, for visualization purposes, the figure shows the translated geometric mean return G-1. Shades of red refers to negative returns and shades of blue to positive ones (see colour bar). }
\label{matrix_sharpe_fee_2}
\end{figure}

In this section, we present the results obtained including transaction fees between $0.1 \%$ and $1 \%$ \cite{wikiFees}. In general, one can not trade a given currency with any given other. Hence, we consider that each day we trade twice: We sell altcoins to buy Bitcoin, and we buy new altcoins using Bitcoin. The mean return obtained between Jan. 2016 and Apr. 2018 is larger than $1$ for all methods, for fees up to $0.2\%$ (see \cref{fees_table}). In this period, Method 3 achieves positive returns for fees up to $1\%$. The returns obtained with a $0.1\%$ (see \cref{matrix_sharpe_fee_1}) and $0.2\%$ (see \cref{matrix_sharpe_fee_2}) fee during arbitrary periods confirm that, in general, one obtains positive gains with our methods if fees are small enough.

\renewcommand{\arraystretch}{1.2}

\begin{table}[htb]
\centering
\begin{tabular}{lrrrrrr}
{} &         no fee &      $0.1\%$ &     $0.2\%$ &      $0.3\%$ &      $0.5\%$ & $1\%$ \\
\hline
Baseline &  1.005 &  1.003 &  1.001 &  0.999 &  0.995 &  0.985 \\
Method 1 &  1.008 &  1.006 &  1.004 &  1.002 &  0.998 &  0.988 \\
Method 2 &  1.005 &  1.003 &  1.001 &  0.999 &  0.995 &  0.985 \\
Method 3 &  1.025 &  1.023 &  1.021 &  1.019 &  1.015 &  1.005 \\
\hline
\end{tabular}
\caption{\textbf{Daily geometric mean return for different transaction fees.} Results are obtained considering the period between Jan. 2016 and Apr. 2018. }
\label{fees_table}
\end{table}

\subsection{Results in USD}
\label{results_usd}
\begin{figure}[htb]
\centering
\includegraphics[width=0.9\textwidth]{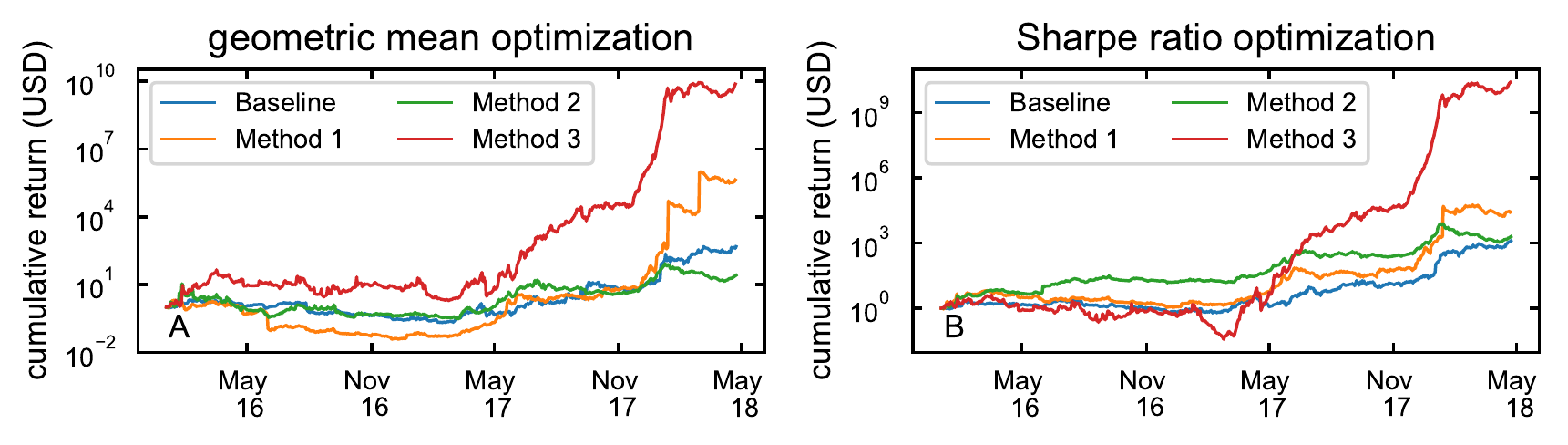}
\caption{\textbf{Cumulative returns in USD.} The cumulative returns obtained under the Sharpe Ratio optimisation (A) and the geometric mean optimisation (B) for the baseline (blue line), Method 1 (orange line), Method 2 (green line) and Method 3 (red line). Analyses are performed considering prices in BTC.}
\label{cumulative_usd}
\end{figure}

In this section, we show results obtained considering prices in USD. The price of Bitcoin in USD has considerably increased in the period considered. Hence, gains in USD (\cref{cumulative_usd}) are higher than those in Bitcoin (\cref{cumulative}). Note that, in \cref{cumulative_usd}, we have made predictions and computed portfolios considering prices in Bitcoin. Then, gains have been converted to USD (without transaction fees). In \cref{table_USD}, we show instead the gains obtained running predictions considering directly all prices in USD. We find that, in most cases, better results are obtained from prices in BTC.

\begin{table}[htb]
\centering
\begin{tabular}{lrr}
\centering
{} &       Geometric mean in USD (from BTC prices) & Geometric mean in USD (from USD prices) \\
\hline
Baseline &  1.0086 & 1.0141 \\
Method1  &  1.0121 & 1.0085 \\
Method2  &  1.0091 & 1.0086 \\
Method3  &  1.0289 & 1.0134 \\
\hline
\end{tabular}
\caption{\textbf{Geometric mean returns in USD.} Results are obtained for the various methods by running the algorithms considering prices in BTC (left column) and USD (right column).}
\label{table_USD}
\end{table}

\subsection{Geometric mean optimisation}
\begin{figure}[th]
\centering
\includegraphics[width=0.7\textwidth]{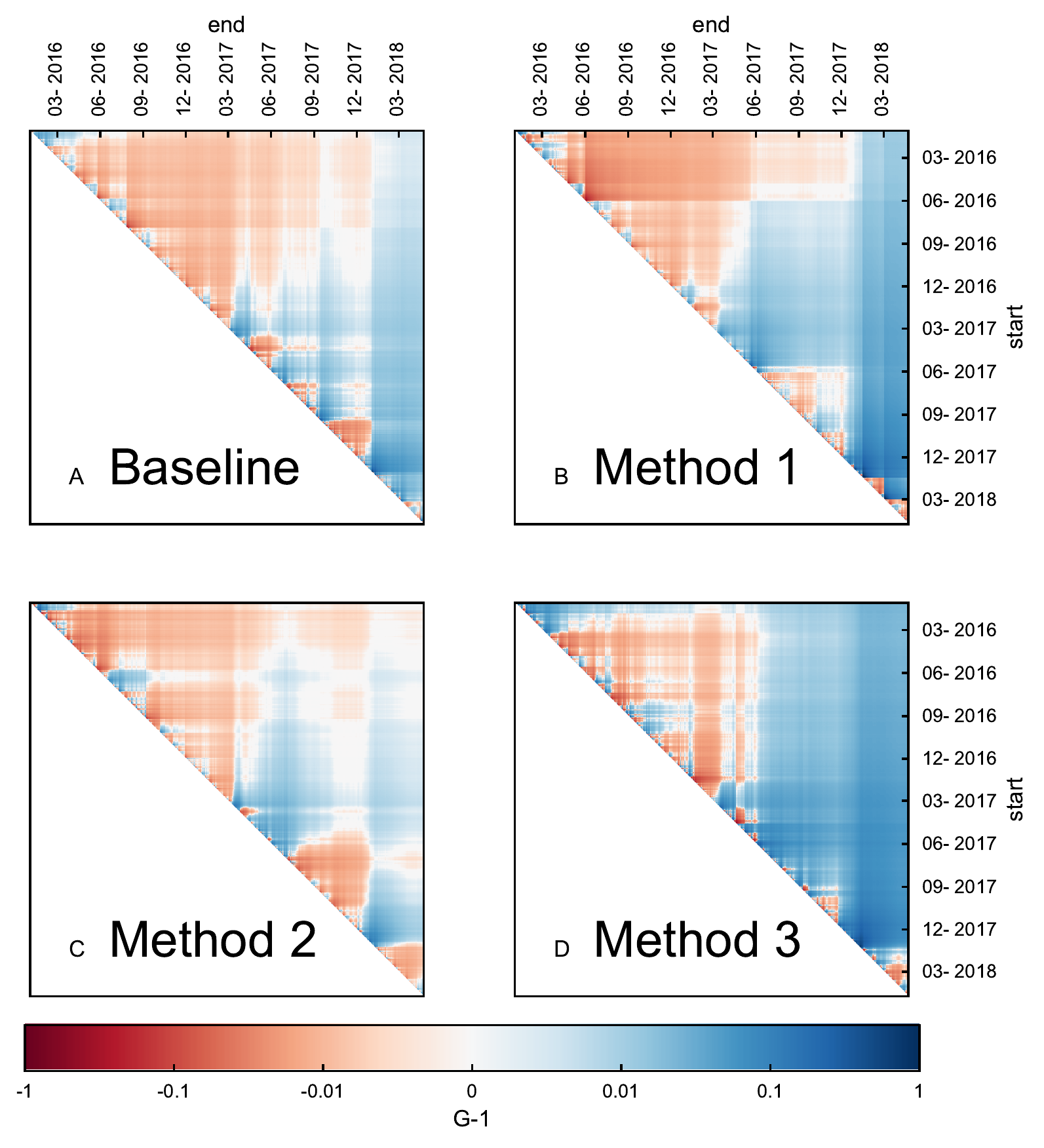}
\caption{\textbf{Geometric mean return obtained within different periods of time.} The geometric mean return computed between time "start" and "end" using the Sharpe ratio optimisation for the baseline (A), Method 1 (B), Method 2 (C) and Method 3 (D). Note that, for visualization purposes, the figure shows the translated geometric mean return G-1. Shades of red refers to negative returns and shades of blue to positive ones (see colour bar). }
\label{matrix_geom}
\end{figure}

In \cref{matrix_geom}, we show the geometric mean return obtained by between two arbitrary points in time under geometric mean return optimisation for the baseline (\cref{matrix_geom}-A), Method 1 (\cref{matrix_geom}-B), Method 2 (\cref{matrix_geom}-C), and Method 3 (\cref{matrix_geom}-D). 

\end{document}